# Effect of Passive flow Control of Bifurcation Phenomenon in Sudden Expansion channel

A. Mishra[1], M. Jithin[2], A. De[1] and M. K. Das[2]

[1]Department of Aerospace Engineering, Indian Institute of Technology Kanpur, Kanpur,208016, India

[2]Department of Mechanical Engineering, Indian Institute of Technology Kanpur, Kanpur, 208016, India

## Abstract

In this article we describe, a numerical simulation has been performed for a channel flow with sudden expansion. Different geometric expansion ratios are considered in the current study such as 1:2, 1:3, and 1:4. The bifurcation phenomenon has been investigated for the baseline case and the critical Reynolds number is controlled using a passive modification in the channel. The geometry of sudden expansion channel is modified by introducing a two-dimensional plate along the center line of the channel at the expansion plane. This is one of the passive control techniques, which has been explored in the present study. The computations indicate that the critical Reynolds number of the symmetry-breaking bifurcation increases as the length of the plate increases. An empirical relation has been developed amongst the critical Reynolds number, length of control plates and expansion ratio ($h/H$). The critical Reynolds number predicted by the empirical relation for different plate lengths are in good agreement with computational results.

**Nomenclature**

| | | | |
|---|---|---|---|
| $u_i$ | Cartesian component of velocity | $Re_c$ | Critical Reynolds number |
| $x_i$ | Cartesian coordinate vector component | $U$ | Velocity in the x direction |
| $v$ | Kinematic viscosity | $h/H$ | Expansion ratio |
| $Re$ | Reynolds number | N-S | Navier-Stokes |

## 1. Introduction

The flow separation on sudden expansion geometries is a well-known phenomenon. Several experiments and theoretical techniques were developed and emphasized in the literature [1-5] to study such phenomenon. The usage of this phenomenon can be seen in a device like the entry regions of fuel cells, reactors, and heat exchangers. The geometry plays a pivotal role, especially in the case of flow through sudden expansion geometry, which finds its applications in burners and cyclone separators [6]. The investigation of such flow brings an insight into separation mechanism in both laminar and turbulent flow regimes It was found that there exists a critical value of the Reynolds number, $Re_c$ such that, when $Re < Re_c$ the flow in the channel is steady, two-dimensional and symmetric, while for $Re > Re_c$ the flow still remains two-dimensional and steady (up to certain $Re$), but the behavior of the flow changes to asymmetric. This phenomenon is known as symmetry-breaking bifurcation. It has also been observed that below this critical Reynolds number both the recirculation regions are of the same size formed on the both side of the channel wall. Further increase in Reynolds number causes one of the recirculation lengths to increase in the expansion of others. The asymmetry in the channel arises due to disturbance generated at the edge of the expansion which amplifies in the shear layer with increase of $Re$. At critical Reynolds number, these flows incur instabilities that may lead to a bifurcation, unsteadiness, and chaos [7-8].

Durst et al. [9] conducted the flow visualization and Laser Doppler Velocimetry (LDV) measurements in a plane-symmetric sudden expansion (1:3) duct. They used air as working fluid and it was found that flow behavior at the downstream of the channel clearly depends on Reynolds number. After a certain Reynolds number, flow changes from symmetric to asymmetric which is subsequently followed by strongly three-dimensional nature of flow even well away from the channel corners. Cherdron et al. [10] conducted experiments on the sudden expansion channels for expansion ratio 1:2 and 1:3 with various aspect ratios (length of the diffusor duct to inlet height) and observed that irrespective of expansion ratios, the critical Reynolds number decreases with the increase in the aspect ratio. The critical Reynolds number for $h/H=$1:2 for varying parameters is found in the range of 120-200. Fearn et al. [11] studied the origin of steady asymmetry flows in a symmetric sudden expansion ratio of 1:3 using the experimental and numerical technique. They reported that the flow becomes time dependent and shows the three-dimensional effect at high Reynolds number. Durst et al. [12] investigated flow in a sudden expansion channel with expansion ratio 1:2 and demonstrated experimentally and numerically that the flow becomes asymmetric above Reynolds number 125. It was also noted that the flow remained laminar upto $Re$=610 and shows slightly three-dimensional effect beyond this $Re$. Alleborn et al. [13] carried out numerical study and linear stability analysis for flow in a channel with symmetric and asymmetric sudden expansion with a range of expansion ratios and compared the results with experimental data. It was reported that a sequence of bifurcation points are encountered when the branch of unstable symmetric solutions beyond the first bifurcation point is traced up to higher Reynolds number. Luo [14] numerically investigated nonlinear flow phenomenon in the 2-D symmetric channel using Lattice-Boltzmann Method (LBM). It was found that symmetry breaking bifurcation takes place at Reynolds number 46.7 for $h/H=$1:3. Drikakis [8] carried out a computational study of laminar incompressible flows in symmetric plane sudden expansion for different expansion ratios including $h/H=$1:2, 1:3 *and* 1:4 in which it was inferred that the critical Reynolds number reduced with increase in the expansion ratio. Battaglia et al. [15] numerically investigated the flow field in the two-dimensional channel on multiple expansion ratios to understand how the channel expansion ratio influences the symmetric and asymmetric solutions and found the critical Reynolds number for expansion ratio of 1:2, 1:3 and 1:4 to be 150, 57 and 35, respectively. Hawa and Rusak [7] carried out bifurcation analysis, linear stability study and direct numerical simulations of two-dimensional, incompressible and laminar flows in the symmetric long channel with the sudden expansion of ratio 1:3 and found that the critical Reynolds number to be 53.8. Fani et al. [16] studied linear stability analysis on sudden expansion channel with ratio 1:3. The sensitivity analysis of the solution was carried out with the structural perturbation of the linearized N-S equations and perturbation of the base flow in order to calculate flow instability. The combined study of two techniques was then used for passive flow control by introducing a small cylinder at the expansion plane to stabilize the unstable symmetric flow in the diffuser duct.

Although, several studies including experimental, numerical and stability analysis have been carried out on bifurcation characteristic of flows in sudden expansion channel; as per the knowledge of the author, only single literature [16] exists which discusses about passive control of bifurcation in sudden expansion flows. Efforts have been made to take a step ahead for a deeper understanding of passive control on sudden expansion flows and its effect on the critical Reynolds number. The objective of this work is to study the two-dimensional flow in sudden expansion channel and control the bifurcation effect using passive control method.

## 2. Simulation Details

### 2.1 Governing equations

The non-dimensional governing equations for viscous Newtonian incompressible flows are given by Eq. (1) and (2)

$$\text{Continuity equation: } \frac{\partial u_i}{\partial xi} = 0 \tag{1}$$

$$\text{Momentum equation: } \frac{\partial u_i}{\partial t} + u_j \frac{\partial u_i}{\partial x_j} = -\frac{1}{\rho} \frac{\partial p}{\partial x_i} + \nu \frac{\partial^2 u_i}{\partial x_i^2} \tag{2}$$

## 2.2 Computational domain and Boundary condition

The 2-D computational domain, shown in Fig. 1, comprises a channel inlet section of height $h$ that expands in to a channel of height $H$, i.e. the expansion ratio is $h/H$. The simulations have been carried out with three expansion ratios ($h/H$=1:2, 1:3 and 1:4). The inlet section of the channel is considered three time of inlet height $h$ and length of diffuser duct is $60h$. The passive control has been done by placing a 2-D plate in the middle of expansion plane which is also shown in Fig. 1.

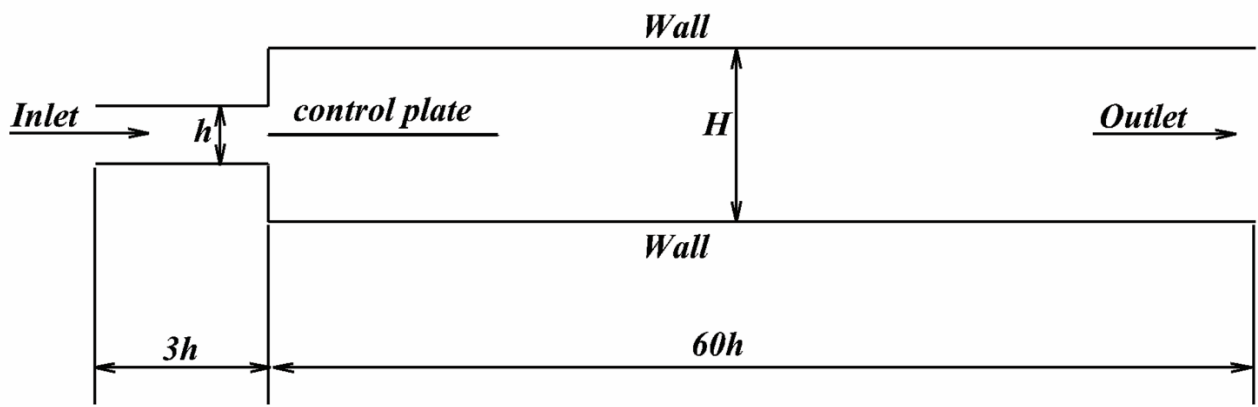

**Fig. 1** Schematics of a symmetric sudden expansion channel with boundary conditions.

A fully developed flow condition is specified at the inlet for velocity, based on channel height and Reynolds number while a zero gradient boundary condition is applied for pressure. At the outlet, convective boundary conditions are specified, whereas no-slip boundary condition is enforced at top and bottom walls.

## 2.3 Numerical Details

The governing equations of fluid flow are solved in OpenFOAM framework [17]. The incompressible solver, pisoFoam is utilized for a solution, based on the finite-volume (FVM) discretization method. Pressure and velocity coupling are handled using PISO (Pressure Implicit with splitting of Operators) algorithm. Second order discretization schemes are used for all the terms in the governing equations. An adjustable variable time integration is used to guarantee the local Courant number less than 1 (CFL<1). Tolerance is set to $10^{-6}$ for all variables to get an accurate solution at each time step.

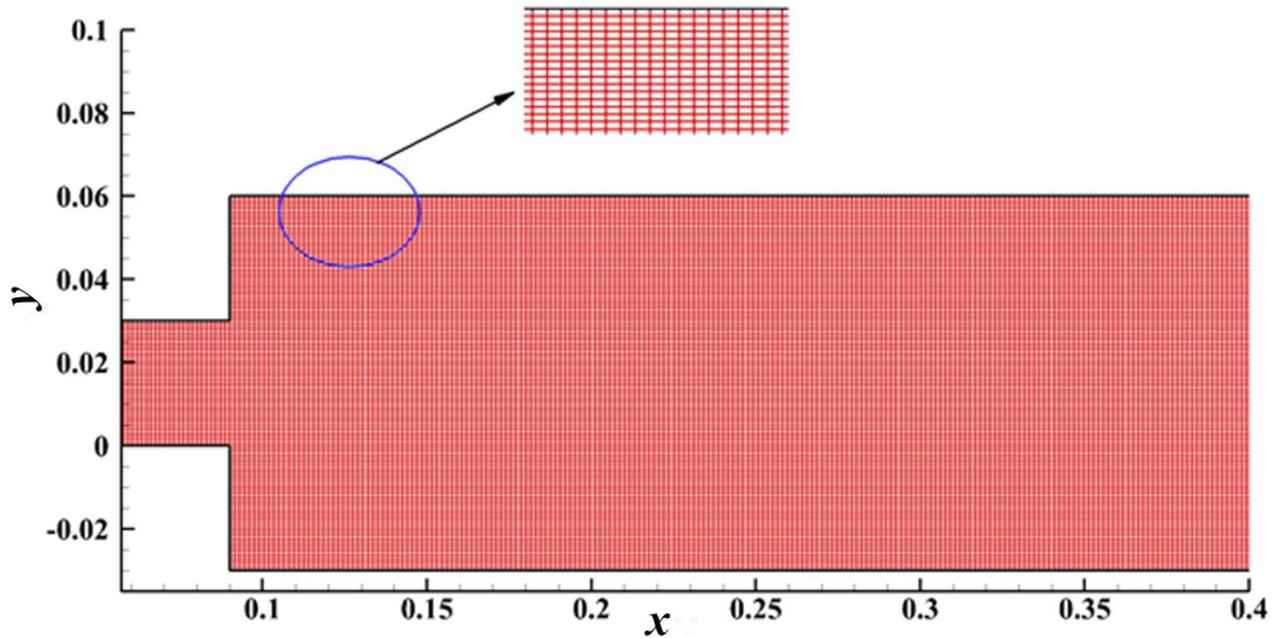

**Fig. 2** 2-D computational grid for sudden expansion channel

Two-dimensional computational grids used in the current simulations are generated using ANSYS® ICEM-CFD [18]. The grid is fully resolved upto $y+=1$ near the walls of the channel. Grid comparison has been performed with three different grids for Reynolds number 40. Finally, simulation has been carried out with a grid of size 141,516 number of the grid cells as shown in Fig. 2.

**3. Grid Independence Test ($h/H=1:3$)**

Grid comparison has been performed with three different grids for Reynolds number 40 as tabulated in Table 1.

**Table 1.** Computational grids for simulation

| Grid Type | Grid Size |
|:---:|:---:|
| Coarse | 141,516 |
| Medium | 322,366 |
| Fine | 668,446 |

It can be observed from Fig. 3 that by increasing the number of grid cells, there seems to have no significant difference in the computed velocity profile as compared to the results obtained by the coarse grid. Hence, the coarse grid has been selected for further analysis.

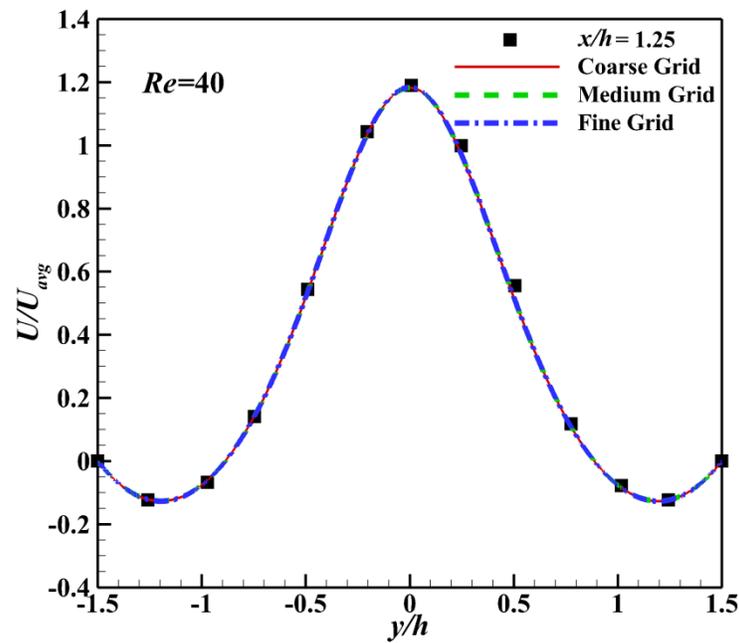

**Fig. 3** Grid independence test for $Re=40$, compared with experimental measurements [11] at $x/h = 1.25$ along the channel. Symbols represent the experimental data and lines denote the predictions.

## 4. Results and Discussion

### 4.1 Baseline expansion channel ($h/H$=1:3)

The velocity profiles of the baseline sudden expansion channel have been compared with the experimental measurements of Fearn et al. [11] at various cross-sections along the channel for Reynolds number 40, 80 and 187, as shown in Fig. 4. The present simulated results provide excellent predictions of the flow fields in all the cases, similar to the predictions reported in the published literature [7, 11, 15 and 19]. It has also been observed through the velocity profile that the flow remains symmetric for $Re$=40 and it becomes asymmetric for $Re$=80 and 187. Furthermore, it can also be concluded that the symmetry breaking bifurcation takes place between $Re$=40 and 80.

To further show the pattern of the flow field, Fig. 5(a) depicts the streamlines plot for $Re$=35, which confirms the symmetric behavior of the flow field at this Reynolds number and the length of the recirculation bubble happens to be the same on both sides of the corner. Whereas, Fig 5(b) exhibits the asymmetric streamlines for $Re$=60, where the bubble length is different on both sides of the corner. Thereafter, a series of simulations for varying $Re$ has been carried out for the baseline sudden expansion channel. Notably, flow remains symmetric up to Reynolds number 53, which can also be witnessed from the bifurcation diagram for baseline case and passive control case as shown in Fig. 6. Similar Navier–Stokes simulations have been carried out in many other studies which are also compared in Table 2. It is worth mentioning that all the published literature show similar behavior including our present study, and in fact, our simulation exhibit quite a good match with the results of Hawa and Rusak [7] and Battaglia et al. [15] while the obtained critical Reynolds number is around 53.

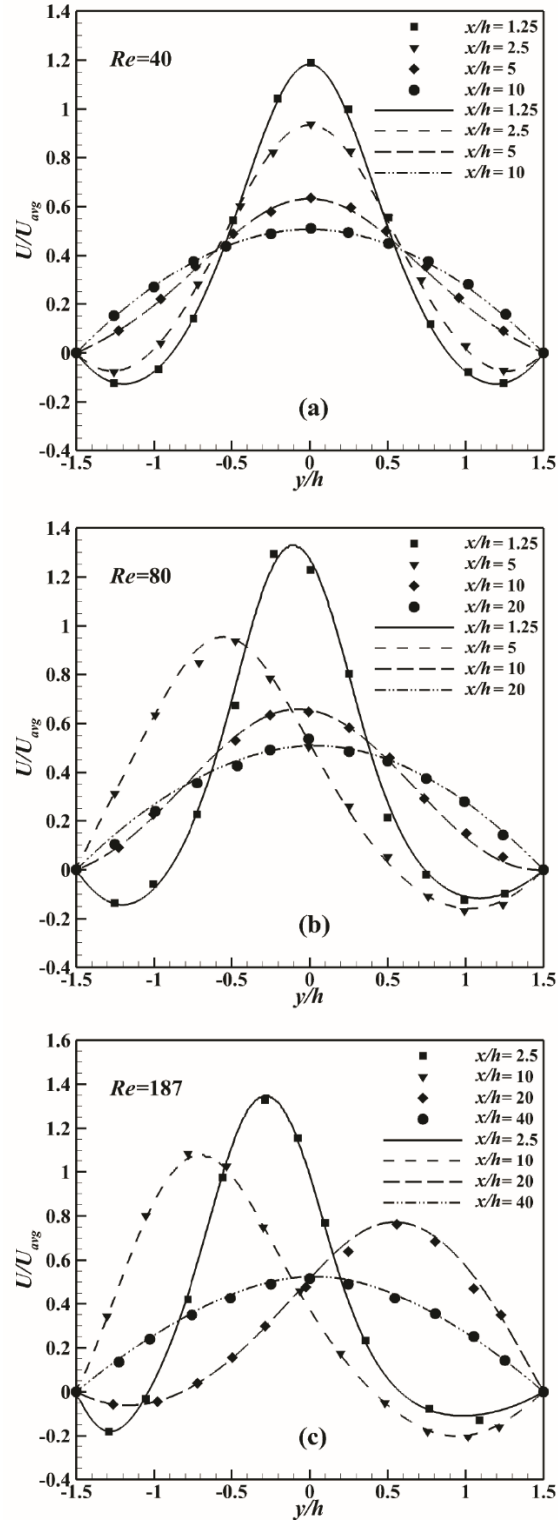

**Fig. 4** Velocity profiles for symmetric sudden expansion, (a) *Re*=40 (b) *Re*=80 (with Hawa and Rusak[7]), and (c) *Re*=187 (with Battaglia et al. [15]). Symbols represent the measurements and lines represent the predictions.

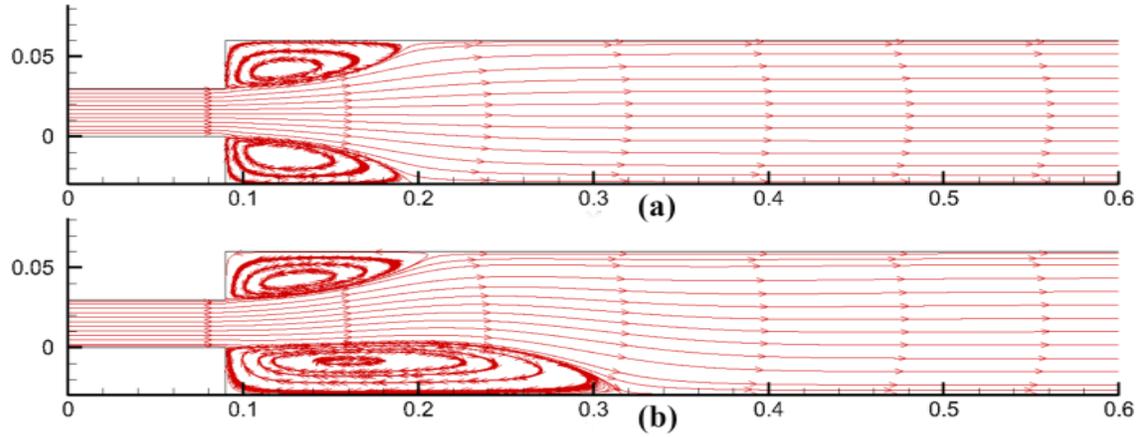

**Fig. 5** Instantaneous streamlines plot of baseline case for **(a)** *Re*=35 and **(b)** *Re*=60

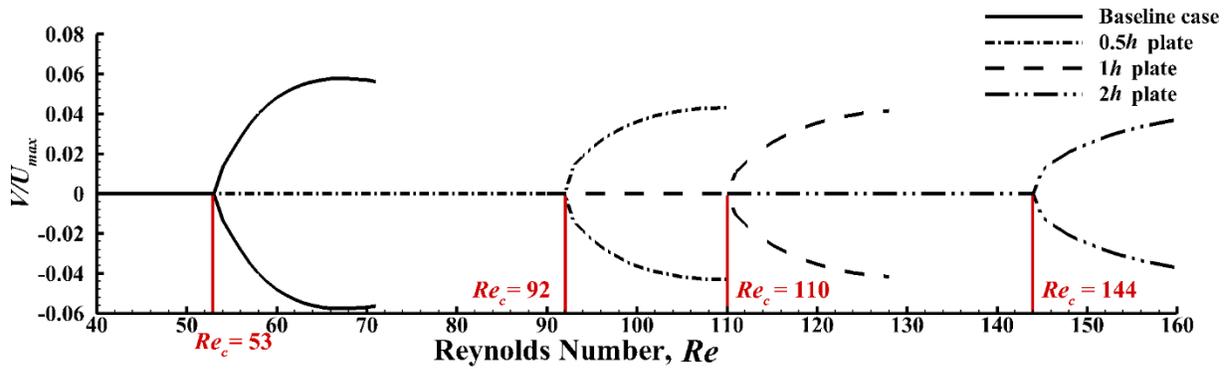

**Fig. 6** Bifurcation plot for baseline and passive control cases. Where, *V* is considered to be the velocity component in the *y* direction at a point on the centerline of the channel, at a distance of 6.38*h*, 6.88*h*, 7.38*h*, 8.38*h* and 10.38*h* from the plane of expansion baseline, 0.5*h*, 1*h*, 2*h* and 4*h* length of control plate respectively. It is same for all the cases studied herein.

**Table 2.** Symmetry-breaking bifurcation for different studies

| Author | Expansion ratio | Method | Critical *Re* |
|---|---|---|---|
| Durst et al. [9] | 1:3 | Experimental | 56 |
| Battaglia et al. [15] | 1:3 | Stability analysis | 53.8 |
| Battaglia et al. [15] | 1:3 | Simulation | 56-57 |
| Hawa and Rusak [7] | 1:3 | Simulation | 53.8 |
| **Present N-S Study** | **1:3** | **Simulation** | **53** |

## 4.2 Passive control with 2-D plate (*h/H* =1:3)

The geometry has been modified by introducing a two-dimensional plate at the central line of the channel in the expansion plane which is known as passive control of the flow. Velocity contour plots for the baseline case and a case with a plate of length *h* is elucidated in Fig. 7 for comparative study. An increase in velocity is seen on both sides of the corner in the diffusion duct due to the control plate which results in the delay of the flow interaction at the center line of the channel. Consequently, there is a significant increase in the critical Reynolds number. The delay in the flow interaction at the center line is also confirmed through Fig. 8 which shows the *y* direction velocity along the centerline of the channel. Velocity in the *y*-direction is zero throughout the domain at pre-critical Reynolds number because of the symmetric recirculation bubble formed at the lower and upper walls but a sinusoidal variation in the velocity profile (Fig. 8) is obtained due to the asymmetry in the flow at post-critical Reynolds number. With the insertion of the passive control plate at the mid of the expansion plane, the *y* velocity profile gets shifted by a length equal to the length of the control plate.

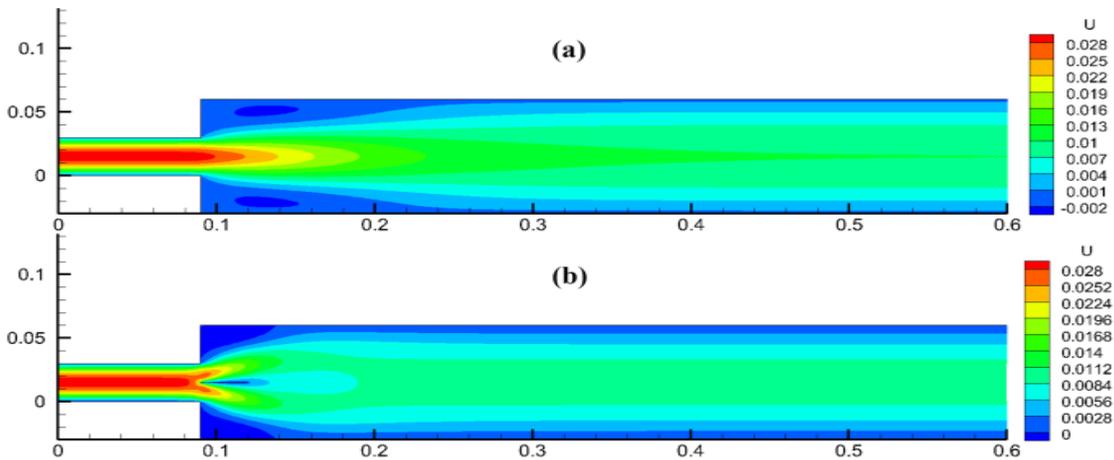

**Fig. 7** Instantaneous velocity contour plots of **(a)** baseline case at *Re*=40 and **(b)** *1h* plate control at *Re*=40

This postpones the interaction of flow along the center line which leads to the increase in the symmetry-breaking bifurcation point. A two-dimensional plate of length *0.5h, h, 2h and 4h* has been utilized as a passive control. The streamline plots for these control plate are shown in Fig. 9.

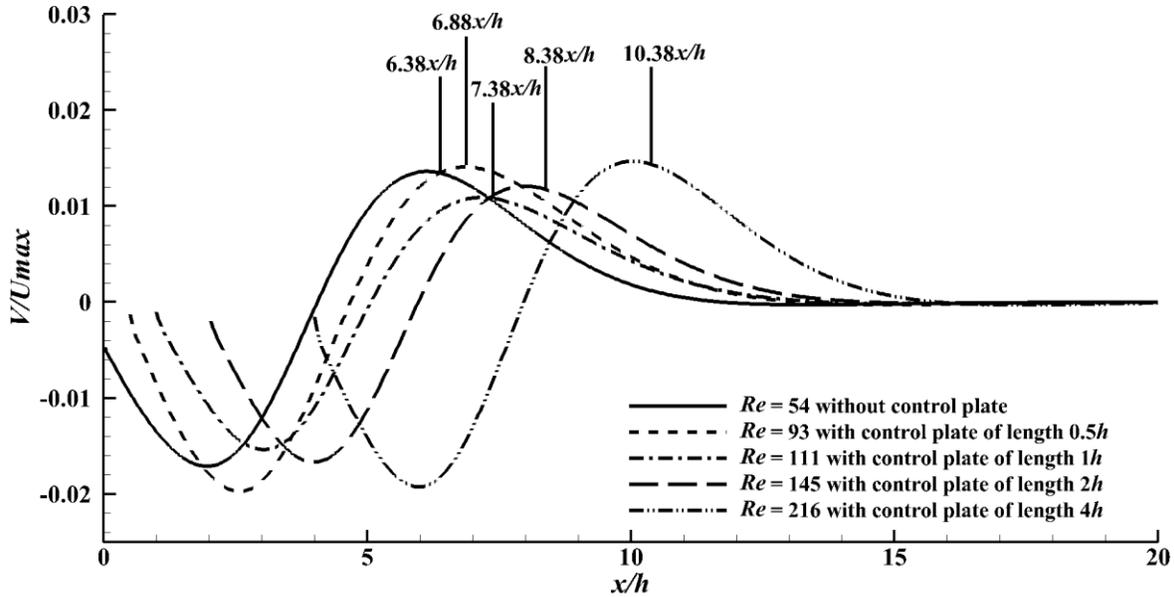

**Fig. 8** The velocity component in the *y*-direction at the centerline of the channel for baseline and passive control cases.

It has been observed through Fig. 9(a) and (b) that the flow is symmetric at *Re*=80 and asymmetric at *Re*=104 for plate length of 0.5*h*. Precisely, the flow bifurcation takes place at *Re*=92, beyond which flow becomes asymmetric for this control length plate, which can also be seen in bifurcation plot, as depicted in Fig.6.

Fig. 9(c) and (d) depict the streamline plots for a *1h* plate case. In this case, the flow appears to be symmetric at *Re*=104 and asymmetric at *Re*=119. It shows that the critical *Re* lies in between this Reynolds number range, and found to be 110 for this case.

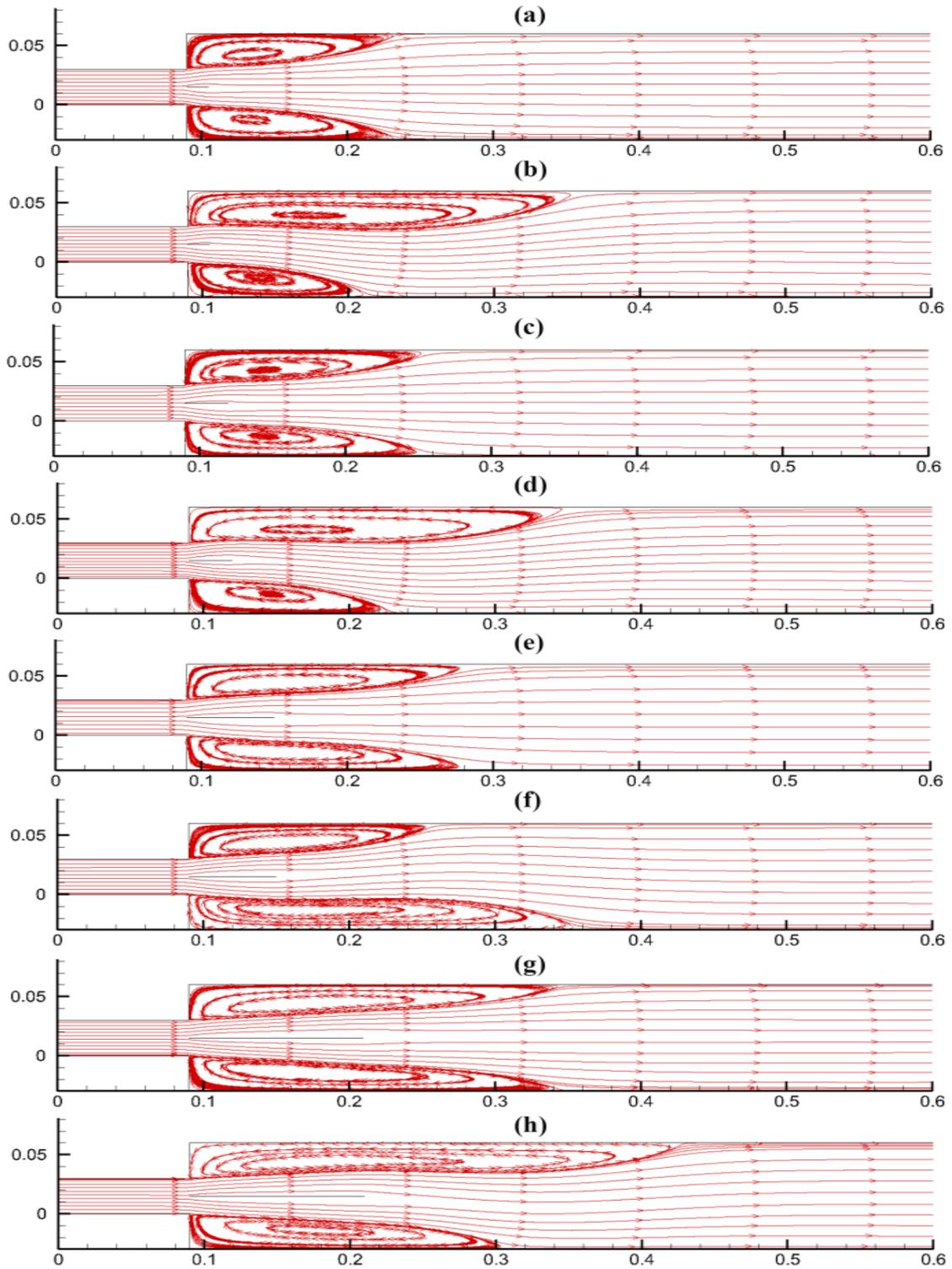

**Fig. 9** Instantaneous streamline plot of passive control with 0.5*h* length plate for **(a)** *Re*=80 and **(b)** *Re*=104, with 1*h* length plate for **(c)** *Re*=104 and **(d)** *Re*=119, with 2*h* length plate for **(e)** *Re*=138 and **(f)** *Re*=150 and with 4*h* length plate for **(g)** *Re*=207 and **(h)** *Re*=220

While comparing different plate lengths in Fig. 9(b) and (c), the flow field is found to be asymmetry for control plate length of *0.5h* while flow remains symmetric for control plate length *1h* at the same Reynolds number, i.e. *Re*=104. This clearly shows that there is an increment on the critical Reynolds as the length of control plate increases and the recirculation length also increases with Reynolds number. The similar trends are also found in the cases of control plate lengths of *2h* and *4h*, shown in Fig. 9(e-h). The observed critical Reynolds numbers for control plate lengths of *2h* and *4h* are 144 and 215, respectively.

**4.3 Bifurcation analysis on *h/H*=1:2 and 1:4**

The study has been extended beyond 1:3 expansion ratio and two more expansion ratios of 1:2 and 1:4 are invoked in the present investigation to have a deeper insight of the sudden expansion flow physics. The simulated results for the baseline cases have been compared with previous studies as shown in Table 3. The results are in close agreement with Battaglia et al. [15].

**Table 3.** Comparison of symmetry-breaking bifurcation with different study without control

| Author | Expansion ratio | Method | Critical *Re* |
|---|---|---|---|
| Battaglia et al. [15] | 1:2 | Simulation | 150-155 |
| Battaglia et al. [15] | 1:2 | Bifurcation Calculations | 143.6 |
| **Present N-S Study** | **1:2** | **Simulation** | **146** |
| Battaglia et al. [15] | 1:4 | Simulation | 35-40 |
| Battaglia et al. [15] | 1:4 | Bifurcation Calculations | 35.8 |
| **Present N-S Study** | **1:4** | **Simulation** | **35** |

The critical Reynolds numbers for the cases with passive control plate are illustrated in Table 4.

The effect of the passive flow control plates is found to be similar as *h/H*=1:3. One more important observation is that the critical Reynolds number increases with lowering of the expansion ratios.

**Table 4.** Critical Reynolds numbers of passive control with plate for 1:2 and 1:4 expansion ratios

| Length of passive control plate | Expansion ratio | Critical *Re* |
|---|---|---|
| 0.5*h* | 1:2 | 227 |
| 1*h* | 1:2 | 276 |
| 2*h* | 1:2 | 371 |
| 0.5*h* | 1:4 | 62 |
| 2*h* | 1:4 | 95 |
| 4*h* | 1:4 | 136 |

### 4.4 Empirical formula

An empirical formula has been developed between critical Reynolds number and passive control plate length based on the simulated results of passive control cases with different plate lengths at different expansion ratios. Fig. 10 exhibits the relationship plot between passive control plate length and critical Reynolds number. The linear regression fit between the plate length and the critical Reynolds number is obtained for expansion ratios and represented in Eq. (3-5) as:

$$\text{Re}_c = 34.67h + 74.67 \quad \text{for } h/H=1:3 \tag{3}$$

$$\text{Re}_c = 96h + 179 \quad \text{for } h/H=1:2 \tag{4}$$

$$\text{Re}_c = 22h + 51 \quad \text{for } h/H=1:4 \tag{5}$$

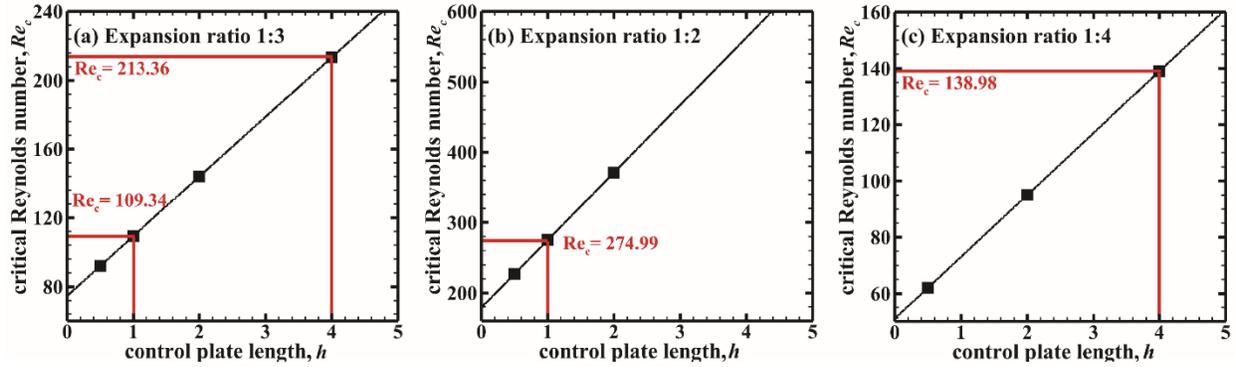

**Fig. 10** Linear empirical Relationship curve between passive control plate lengths and respective critical Reynolds numbers. The critical Reynolds numbers shown in figures are evaluated from the empirical formula.

Now, the relationship between expansion ratio ($h/H$), slope of linear variation and intercept is established to come up with the general form of the empirical relations. The values of expansion ratio ($h/H$), slope of linear relations of expansion ratio and intercept for each linear relation is shown in Table 5. The variables shown in Table 5 are considered to arrive at a generalized empirical relationship. Here the variable expansion ratio ($h/H$) was first varied with the slope of linear relations and then with the intercept as displayed in Fig. 11.

**Table 5.** Parameters for empirical relationship

| Expansion ratio ($h/H$) | Slope of linear variations | Intercept |
|---|---|---|
| 1:2 | 96 | 179 |
| 1:3 | 34.67 | 74.67 |
| 1:4 | 22 | 52 |

It appears that a quadratic regression fit between $h/H$ and slope is obtained as shown in Fig. 11(a) and exhibited in Eq. (7). Fig. 11(b) depicts that the similar quadratic expression is also found for the $h/H$ and intercept variations as described in Eq. (8). Combining all of these, the general

formula is represented in Eq. (6), (7) and (8)

$$Re_c = f(h/H) \cdot h + g(h/H) \tag{6}$$

$$f(h/H) = 55.93 - 351.6(h/H) + 863.47(h/H)^2 \tag{7}$$

$$g(h/H) = 93.88 - 513.29(h/H) + 1367.05(h/H)^2 \tag{8}$$

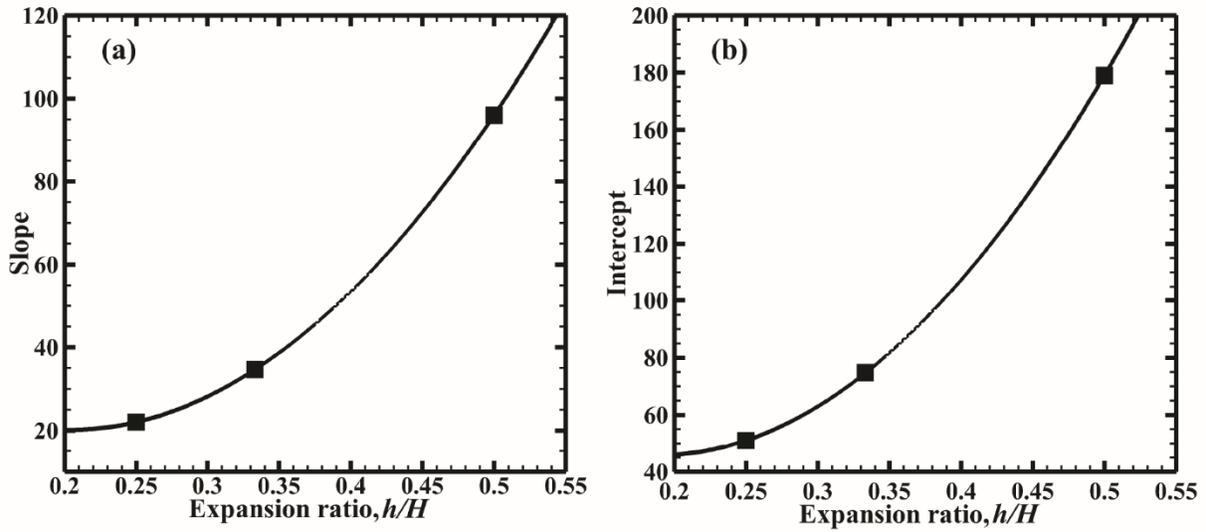

**Fig. 11** (**a**) Variation of slope of linear relations with expansion ratio, $h/H$ (**b**) Variation of intercept with expansion ratio

**Table 6.** Comparison of critical *Re* with empirical relationship

| Expansion ratio | Control plate length | Critical *Re* through simulation | Critical *Re* through empirical formula | Error (in %) |
|---|---|---|---|---|
| 1:3 | 1$h$ | 110 | 109.34 | 0.6 |
| 1:3 | 4$h$ | 215 | 213.36 | 0.76 |
| 1:2 | 1$h$ | 276 | 274.99 | 0.36 |
| 1:4 | 4$h$ | 136 | 138.98 | 2.19 |

## 4.4 Verification of empirical formula

The empirical relation has been verified with different lengths of control plate and results are tabulated in Table 6. The symmetry-breaking bifurcation points through empirical formula are also shown in Fig 10. The prediction of critical Reynolds number through empirical relation is closely matched with the simulated critical Reynolds number for all the cases. It can be observed that maximum error in critical Reynolds number through empirical relation is around 2% only.

## 5. Conclusions

The sudden expansion flow has been investigated for baseline case and passive control case with multiple expansion ratios. The numerical results are compared with the experimental measurements of the axial velocity field at various cross sections for $Re$=40, 80 and 187 in the baseline case with expansion ratio $h/H$=1:3. Simulations provide a good prediction of flow fields in all the three cases, while the symmetry-breaking bifurcation for $h/H$=1:3 has been observed at $Re$=53 for the baseline case. Similar analyses have also been extended for the expansion ratios of 1:2 and 1:4 to have a better understanding of the sudden expansion flow behaviour. It is found that the critical Reynolds number for the baseline cases are 146 ($h/H$=1:2) and 35 ($h/H$=1:4), respectively. Channel has been modified with passive control as a 2D plate and the passive control significantly increases the critical Reynolds number. An empirical relationship among the plate length, critical Reynolds number and expansion ratio ($h/H$) has been developed for passive flow control and the critical Reynolds number predicted by using this formula is in coherence with the simulated data of passive control of plate lengths. The results show that prediction of the critical Reynolds number through empirical formula has around 2% error as compared to simulated data.

**Acknowledgements**

Simulations are carried out on the computers provided by the Indian Institute of Technology Kanpur (IITK) (www.iitk.ac.in/cc) and the manuscript preparation as well as data analysis has been carried out using the resources available at IITK. This support is gratefully acknowledged.